\documentclass{report}

\usepackage{epsfig}

\renewcommand{\abstract}[1]{{ \footnotesize \noindent {\bf Abstract} #1 \\}}

\renewcommand{\author}[1]{\subsubsection*{#1}}
\newcommand{\address}[1]{\subsubsection*{\it#1}}

\newcommand{\1}{\ensuremath{^{-1}}}

\newcommand{\3}{\ensuremath{^{-3}}}
\newcommand{\PC}{\ensuremath{\mathrm{\,pc}}}
\newcommand{\KPC}{\ensuremath{\mathrm{\,kpc}}}
\newcommand{\MPC}{\ensuremath{\mathrm{\,Mpc}}}
\newcommand{\KM}{\ensuremath{\mathrm{\,km}}}
\newcommand{\CM}{\ensuremath{\mathrm{\,cm}}}
\newcommand{\SEC}{\ensuremath{\mathrm{\,s}}}
\newcommand{\YR}{\ensuremath{\mathrm{\,yr}}}
\newcommand{\ERG}{\ensuremath{\mathrm{\,erg}}}

\newcommand{\Msun}{\ensuremath{\,\mathrm{M}_\odot}}
\newcommand{\Rsun}{\ensuremath{\,\mathrm{R}_\odot}}
\newcommand{\Lsun}{\ensuremath{\,\mathrm{L}_\odot}}

\newcommand{\xten}[1]{\ensuremath{\times 10^{#1}}}
\newcommand{\ten}[1]{\ensuremath{10^{#1}}}
\newcommand{\farcs}{\ensuremath{.\!"}}

\newcommand{\kms}{\KM\SEC\1}
\newcommand{\BH}{\ensuremath{\mathrm{BH}}}
\newcommand{\BHs}{\ensuremath{\mathrm{BHs}}}
\newcommand{\Mdot}{\ensuremath{\dot{M}}}
\newcommand{\MBH}{\ensuremath{M_\mathrm{BH}}}
\newcommand{\RBLR}{\ensuremath{R_\mathrm{BLR}}}
\newcommand{\LEdd}{\ensuremath{L_\mathrm{Edd}}}

\newcommand{\Mbulge}{\ensuremath{M_\mathrm{Bulge}}}
\newcommand{\Lbulge}{\ensuremath{L_\mathrm{Bulge}}}
\newcommand{\rhobh}{\ensuremath{\rho_\mathrm{BH}}}
\newcommand{\rbh}{\ensuremath{r_\mathrm{BH}}}
\newcommand{\thbh}{\ensuremath{\theta_\mathrm{BH}}}
\newcommand{\sigmastar}{\ensuremath{\sigma_\star}}

\newcommand{\Sgra}{\ensuremath{\mathrm{SgrA}^\star}}
\newcommand{\aap}{A\&A}
\newcommand{\aj}{AJ}
\newcommand{\apj}{ApJ}
\newcommand{\apjl}{ApJ}
\newcommand{\araa}{ARA\&A}
\newcommand{\mnras}{MNRAS}
\newcommand{\nat}{Nature}
\newcommand{\pasp}{PASP}

\begin{document}

\chapter*{Black Holes in Galactic Nuclei: the Promise and the Facts}

\author{Alessandro Marconi}


\address{Osservatorio Astrofisico di Arcetri,
Largo E. Fermi 5, I-50125 Firenze, Italy}

\abstract{It has long been suspected that Active Galactic Nuclei are powered by
accretion of matter onto massive black holes and this belief implies their
presence in the nuclei of most nearby galaxies as "relics" of past activity.
Just
a few years ago this was considered a paradigm but, recently, new ground-based
and Hubble Space Telescope observations are producing a breakthrough in our
knowledge on massive black holes.  I will review the evidence for the
existence of black holes in galactic nuclei and how their presence is related
to host galaxy properties and AGN activity.}

\section{A Brief Historical Introduction}

In November 1783 John Michell presented to the Royal Society his idea of a {\em
dark star}, a star so massive that the escape velocity from its surface is
larger than the speed of light. Combining the corpuscolary theory of light with
Newton's theory of gravitation, he found that a star with the same density as
the Sun but escape velocity equal to $c$ would have radius $R=486\Rsun$ and
mass $M=1.2\xten{8}\Msun$.  Michell also pointed out that, although dark stars
are invisible, their presence could be inferred from the motion of other
luminous bodies orbiting around them.  Similar ideas were independently
presented in 1796 by Laplace in his {\it "Exposition du syst\`eme du monde"}.

In 1916 Karl Schwarzschild presented his exact solution of Einstein field
equations deriving the well-known {\em Schwarzschild radius} which Michell had
{\it exactly} determined although starting from {\it wrong} assumptions.  The
term {\em black hole}, which is now commonly used, was not coined until 1967 by
John Wheeler and the first observational evidence for the existence of a black
hole was given in the early 70s by the observations of the binary X-ray source
Cygnus X-1.  The discovery of quasars 
with their enormous energy
output from small volumes of space suggested that they were powered by
accretion of matter onto very massive black holes residing in galactic nuclei
\cite{lynden}.  The first observational evidence was found in the galaxy M87
whose nucleus seemed to host a 5\xten{9}\Msun\ black hole \cite{sarg}.
The supermassive black holes (hereafter \BHs) hosted in galactic nuclei, with
masses in the range $\ten{6}-\ten{10}\Msun$, are the topic of this review.
Observational evidences for the existence of \BHs\ in galactic nuclei up to the
early '90s are summarized in a review by Kormendy \& Richstone
\cite{kr}.  At that time only a handful of \BHs\ were known from ground-based
and early HST observations.

\section{Observational Evidences}

There are several reasons why \BHs\ should be present in the nuclei
of active galaxies (e.g. \cite{kr}) but why should \BHs\ be present in normal galaxies?
AGNs
are powered by mass accretion onto a \BH\ and were more numerous and
powerful in the past ($z\simeq 2-3$). Thus one expects that a significant
fraction of local luminous galaxies should host black holes of mass
$10^6-10^{10}$\Msun, relics of past activity (see \S\ \ref{sec:agn&bh}).
For example, a quasar emitting $L=\ten{12}\Lsun$ is powered by 
an accretion rate $\Mdot=L/\epsilon\, c^2\simeq 0.7\Msun/\YR$
(with efficiency $\epsilon=0.1$) onto a \BH.  If the activity lasts for
$\ten{8}\YR$, it will increase the \BH\ mass by $\sim 7\xten{7}\Msun$.

The existence of a \BH\ can be inferred by the gravitational effects on the
surrounding gas or stars, as foreseen by Michell.  In principle, one measures the
velocity field of gas and/or stars in the circumnuclear region of a galaxy and
derives the gravitational potential $\phi$ required to sustain the observed
motions. If the gravitational potential due to the luminous mass
cannot account for $\phi$ then an additional component $\phi_\mathrm{BH}$ is
required to explain the observed motions. If it is spatially "unresolved" at
the observational limit it is called a {\em Massive Dark Object} (MDO) and is a
\BH\ candidate. One can then easily determine \MBH\  
($\phi_\mathrm{BH}(r)=-G\MBH/r$).

{\noindent\bf The two most relevant cases.}
The closest galactic nucleus hosting a \BH\ is our galactic center and this
currently represents the best case for a \BH.  With ground-based
high-spatial-resolution observations (e.g. speckle interferometry, adaptive
optics) it has been possible to measure stellar positions with high accuracies
($\pm 1-5$mas) thus detecting their proper motions. Combining these proper
motions with
spectroscopic measurements, the velocity vector $\vec{v}$ of many stars around
\Sgra\ (the radio source identified with the center of our galaxy) has been directly measured; typical velocities are of the order of a few
100\kms\ with accuracies of $\pm 20-30 \kms$ \cite{genzel00,ghez00,eckart01}.
Ghez et al.\ \cite{ghez00}, using adaptive optics at Keck, have been able to
trace curved stellar orbits, a clear indication of acceleration. Within the
errors, acceleration vectors all intersect at the location of \Sgra. 
All the available data on \Sgra\ have been analyzed in detail by 
Genzel et al.\ \cite{genzel00}
and the main results can be summarized as follows: star
motions can be explained only with a
compact ($\rho_\mathrm{BH} \ge 10^{12.6}\Msun\PC\3$) dark ($M/L>100
\Msun/\Lsun$) mass concentration. Since any dark cluster would have a lifetime
less than $\sim 10^8$yr (see also \cite{maoz98}), too short with respect to the
Hubble time, the data show the presence of a \BH\ with mass
$\MBH=2.6-3.3
\xten{6}\Msun$.  An anisotropy independent estimate of the distance of the
Galactic center is $D=7.8-8.2\KPC$ fully consistent with previous
estimates based on independent methods.
A recent review by Melia \& Falcke presents
the latest observational results and physical interpretation \cite{m&f}. 

The second best case for a \BH\ is in the nucleus of the nearby spiral galaxy NGC 4258, where
high spatial resolution VLBA spectroscopic observations of the H$_2$O maser
emission have shown the presence
of high velocity maser spots. Their velocity field and location in the plane
of the sky
indicate that they are part of a thin, warped disk circularly rotating around
the galactic nucleus with Keplerian velocities. The
velocity field suggests that there is a dark mass concentration of
3.9\xten{7}\Msun\ within 0.14\PC, yielding a mass density of
$\rhobh>4\xten{9}\Msun\PC\3$ \cite{miyoshi}.  
Such an object can only be a black hole because all star clusters with that
density would undergo collapse within a few \ten{8}\YR\ \cite{maoz98}.  The
detection of the maser proper motions, with indications of acceleration, has
allowed a distance estimate accurate to $\simeq 4\%$, similarly to the Galactic
center case. The review by Moran, Greenhill \& Herrnstein \cite{moran} presents
more details and a summary of \BH\ detections with water masers.

The Galactic center and NGC 4258 are really "textbook" cases. Nowadays, proper
motions can be measured only in our Galactic Center, and galaxies with "well
behaving" maser disks are extremely rare.  Out of $\sim 700$ galaxies observed,
22 H$_2$O masers have been detected and only 6 have a "disk" structure which
allows a determination of the \BH\ mass. NGC 4258 is still the most convincing
case \cite{moran}.

{\noindent\bf Results from the Hubble Space Telescope.}
In general, what one can measure is not the velocity vector $\vec{v}$ of a
single star or gas cloud but the overall distribution $f(v)$ of the velocity
components along the line of sight. Furthermore, $f(v)$ is a volume average
over a column elongated along the line of sight with the base set by the spatial
resolution of the observations.
Thus one must deal
with
2-dimensional information of a 3-dimensional structure: detecting a
\BH\ and measuring its mass is not as "simple" and "straightforward" as 
for the Galactic Center and NGC 4258.  I will not discuss
here the detailed methods to measure \BH\ masses with stellar dynamics or gas
kinematics.  The reader can refer to \cite{marel98,magorrian,bower01} for a
detailed description of \BH\ mass measurements with stellar dynamics
and to \cite{macchetto97,marconi01a,barth} for gas kinematics.
In any case, to detect \BHs\ one needs spectral information at the highest
possible angular resolution in order to spatially resolve the \BH\ sphere of
influence \cite{rbh} where the \BH\ dominates over
the galactic potential. The radius \rbh\ of the \BH\ sphere of influence is
\begin{equation}
\rbh = \frac{G\MBH}{\sigma_*^2} \simeq 4.3
\left( \frac{\MBH}{\ten{7}\Msun} \right)
\left( \frac{100\kms}{\sigmastar} \right)^2 \PC 
\end{equation}
where $\sigma_*$ is the velocity dispersion of the stars in the nuclear region.
This can be translated to an angular size in the plane of the sky, 
\begin{equation}
\thbh
\simeq 0.1 \left( \frac{\MBH}{\ten{7}\Msun} \right)
\left( \frac{100\kms}{\sigmastar} \right)^2
\left( \frac{10\MPC}{D} \right) arcsec
\end{equation}
where $D$ is the galaxy distance.
The small values of \thbh\ obtained for typical \MBH, $\sigma_*$
and $D$ values explains why, up to the early 90s, there have been few \BH\
detections from the ground.  The recent breakthrough due to HST is a result of
its high spatial resolution which is almost an order of magnitude better than
from the ground.

I will now briefly outline a few significant cases for \BHs\ from HST
observations.  The first one is that of M87, the giant elliptical galaxy in
Virgo with a radio/optical jet. Sargent et al.\ \cite{sarg} made the first
claim for the presence of a \BH\ with $\sim 5\xten{9}\Msun$ and \cite{harms}
measured high velocities with HST/FOS in the nuclear gas disk thus
strengthening the case for a \BH. The case for the presence of a \BH\ has been
settled with FOC longslit spectra, the first ones obtained from HST: careful
modeling taking into account instrumental effects gave
$\MBH=(3.2\pm0.9)\xten{9}\Msun$ \cite{macchetto97}.  The first STIS detection
of a \BH\ is that in M84  where gas kinematics gave
$\MBH=(1.5\pm0.9)\xten{9}\Msun$ \cite{bower97}.  Using stellar dynamics,
van der Marel et al.\
\cite{marel97} combined HST/FOS and ground based data of the nuclear
region of M32. They detected a dark mass concentration
$\MBH=(3.4\pm1.6)\xten{6}\Msun$ confined within a region of 0.3pc across.  This
was the first detection of a \BH\ in a quiescent galaxy. Since then there have
been many \BH\ detections with HST/STIS stellar dynamical or gas kinematical
studies and the most notable ones are certainly that of NGC 1023 \cite{bower01}
and NGC 3245 \cite{barth}, respectively.

The most secure \BH\ detections up to March 2001 are summarized by Kormendy \&
Gebhardt \cite{kg}. It is immediately clear from their table that most of the \BH\
detections are in E-S0 galaxies (29 out of 36).  In order to fill this gap we
(Axon, Marconi et al.) are just completing an HST/STIS survey of a sample of 54
Sb, SBb, Sc and SBc galaxies \cite{marconi01b}.  Preliminary results from this
survey include the case of NGC 4041 (Marconi et al.\ 2002, in prep), where we have
set a limit of $<\ten{6}\Msun$ to the mass of the \BH, significantly
lower than expected from the  \MBH-\Lbulge\ correlation
(see \S\ \ref{sec:BHgal}) and that of
NGC 4258 (Axon et al.\ 2002, in prep). In the latter case the \BH\ mass estimate
from HST/STIS spectroscopy agrees with the H$_2$O maser estimate thus
confirming the validity of gas kinematical measurements.

{\noindent\bf Ground Based Observations.}
HST has two fundamental limitations: its size, 2.5m, and its lack of a near-IR
longslit spectroscopic facility and both factors do not allow observations of
faint or obscured objects. Eight-meter-class ground-based telescopes with good seeing
and adaptive-optics can overcome these limitations. An example is given by the
detection of a \BH\ in Centaurus A, a famous radio galaxy whose nucleus is
obscured by at least $A_V\sim 7$mag. VLT/ISAAC Pa$\beta$ (1.28$\mu$m)
spectroscopy, with 0\farcs5 seeing, of the nuclear gas disk  has shown the
presence of a \BH\ with $\MBH\sim 2\xten{8}\Msun$ \cite{marconi01a}.
Similarly, Keck Pa$\alpha$ spectroscopy have revealed a \BH\ in Cygnus A with
$\MBH\sim 3\xten{9}\Msun$ (Tadhunter et al.\ 2002, in prep).

{\noindent\bf Going Farther?}
The high spatial resolution required limits \BH\ searches to nearby ($D<100$ \MPC)
objects. How is it possible to go farther? A possibility is offered by
reverberation mapping of broad emission lines in Seyfert 1 galaxies and 
quasars. The time lag, $\tau_\mathrm{BLR}$, between the continuum light
curve and that of a broad line (e.g. H$\alpha$)
is interpreted as light travel time between the compact
continuum source and the more extended {\em Broad Line Region} (BLR): thus $\RBLR=c\tau_\mathrm{BLR}$ 
is an average BLR radius.
Assuming virialized motions of the BLR clouds, 
one can combine $\RBLR$ with the Full Width at Half Maximum (FWHM) of the line and obtain 
$\MBH \simeq 1.45\xten{5}\Msun (c\tau_\mathrm{BLR}/\mathrm{1\,lt\,day})
(FWHM/1000 \kms)^2$ \cite{ho99, wandel99, kaspi00}.
The virial assumption has been
directly tested in the case of NGC 5548 but many
systematic uncertainties are present in this kind of estimate (see \cite{krolik} and references therein).
However, a
strong support for the reliability of this method comes from the agreement 
of \BH\ estimates from reverberation
mapping with the $\MBH-\sigma$ correlation (\S\ \ref{sec:BHgal}).

Another more indirect method uses the FWHM of the broad line (H$\beta$)
combined with the continuum luminosity.  The size of the BLR, estimated from
the correlation between \RBLR\ and the monochromatic luminosity $L_\lambda$ at
5100\AA\ \cite{kaspi00}, is combined with the FWHM of the H$\beta$ line to
derive the \BH\ mass.
In \cite{laor98,mclure01} and references therein the
reader can find applications of this method which seems to provide
reliable \BH\ mass estimates. Note however that the \RBLR-$L_\lambda$(5100\AA)
correlation is not tight at all (see Fig.\ 6 in \cite{kaspi00})!

{\noindent\bf Are they really Black Holes?}
What I have called 
\BHs\ are really massive dark objects.  The
possibility that an MDO is not a cluster of dark objects (e.g. stellar mass
black holes, neutron stars etc.) can be safely ruled out for the
Galactic Center, NGC 4258 \cite{maoz98} and probably M32 \cite{marel97}
on the basis of the short lifetimes of such clusters.
In the two former cases 
observations have probed 
down to $\sim 4\xten{4}$ Schwarzschild radii, still far from the 
General Relativity regime. The
definitive proof that these MDOs are really \BHs\ would be the detection of
relativistic motions close to the event horizon.
This seems to have happened in
the case of the galaxy MCG-6-30-15, where the broad red wing of the K$\alpha$ Fe
line at $\sim 6$keV has been interpreted as due to relativistic effects close to the
event horizon \cite{tanaka95}. 
New XMM observations \cite{wilms01} have confirmed these results and presented
evidences for a rotating black hole and magnetic extraction of \BH\ spin energy
as in the Blandford-Znajek mechanism. Though model dependent, these are very
exciting news and, indeed, the line profile of the K$\alpha$ Fe line is a
unique probe of the region close to the event horizon, in principle allowing
one to distinguish between a Schwarzschild or a Kerr \BH\ \cite{fabian00}.

\begin{figure}[t!]
\centering
\epsfig{file=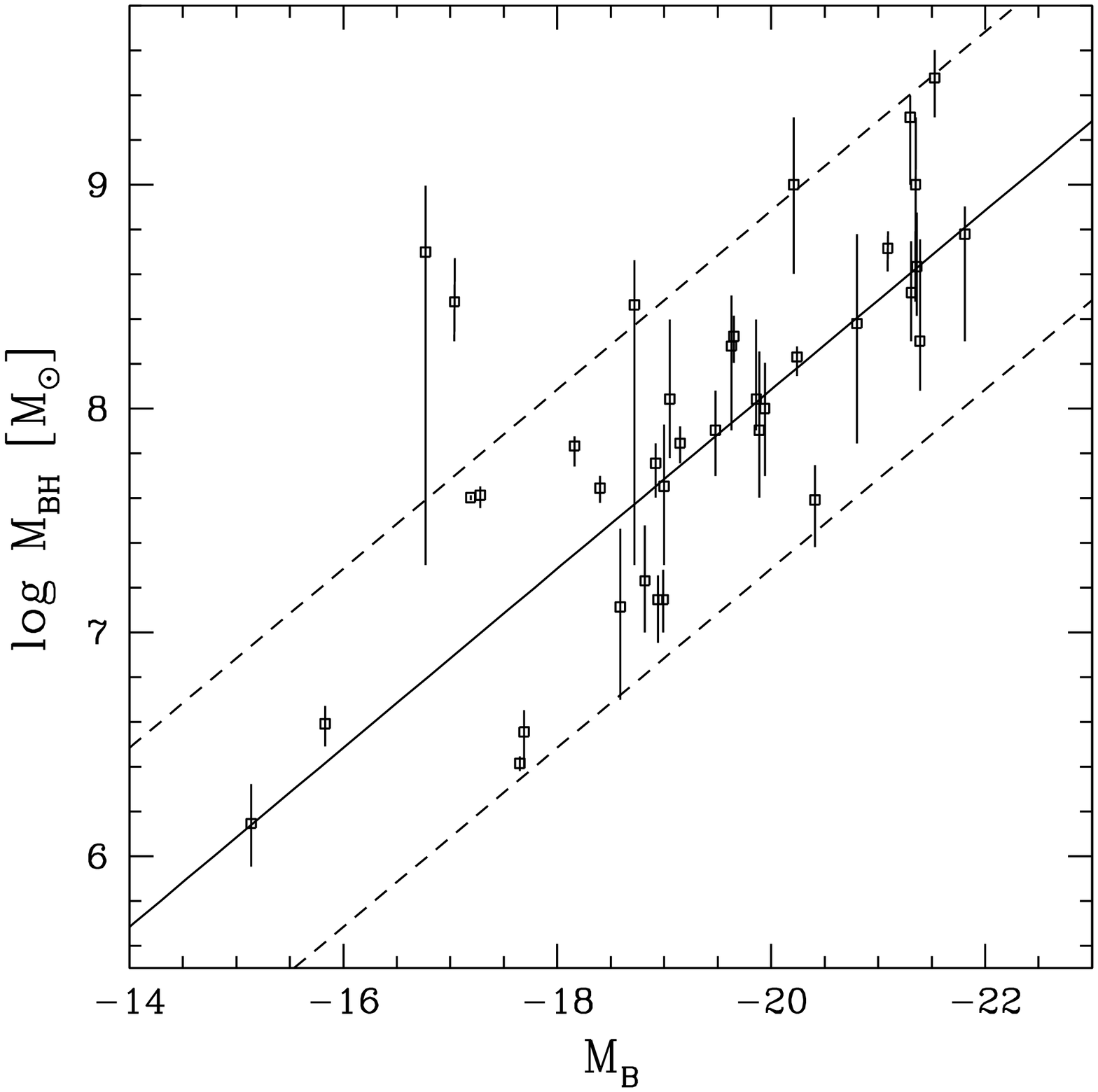, width=0.49\linewidth}
\epsfig{file=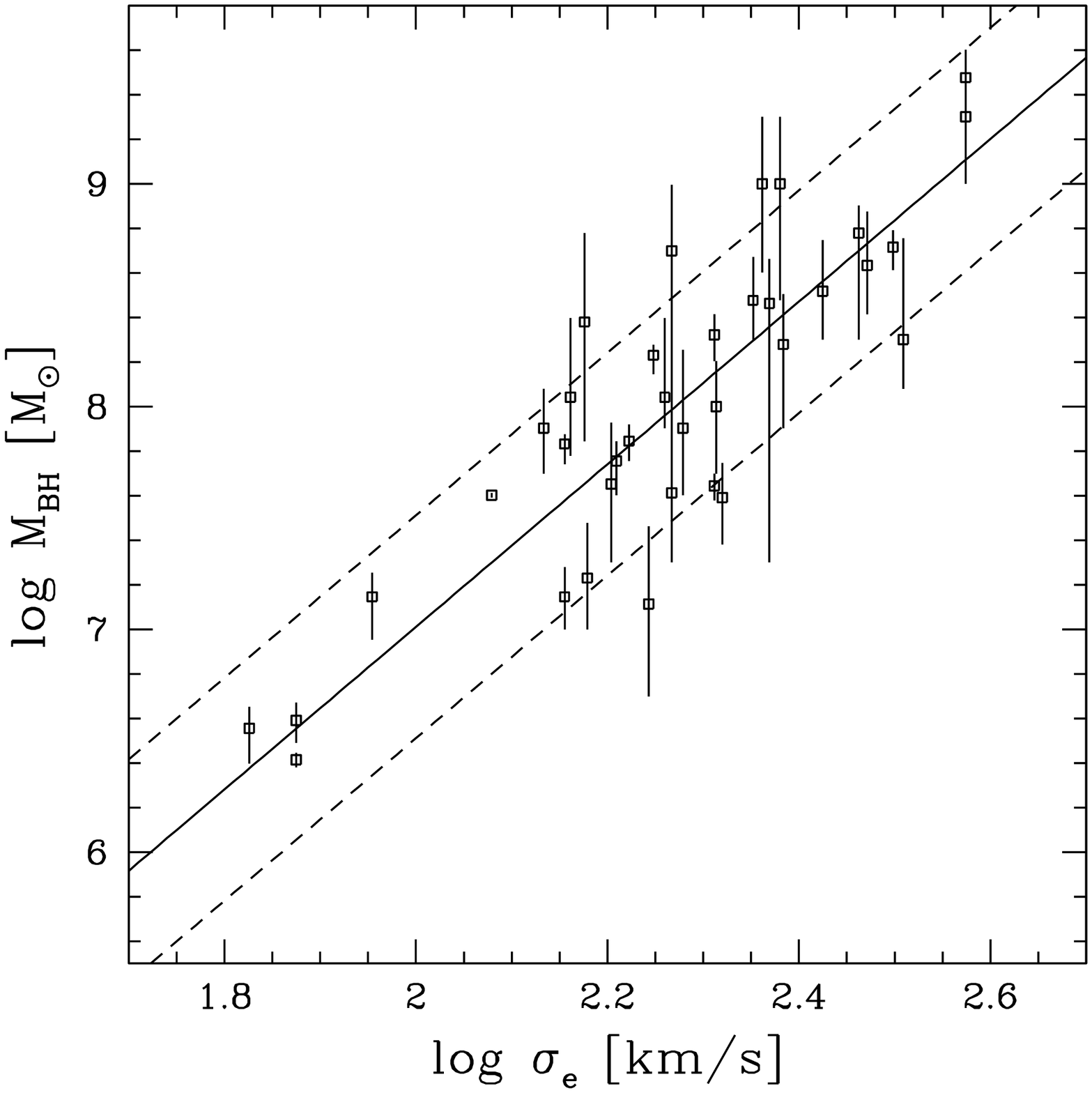, width=0.49\linewidth}
\caption{(a) \label{fig:BHlum}\MBH-\Lbulge\ correlation.
$M_B$ is the absolute magnitude of the bulge in the B band. The dashed lines
represent a $\pm 0.8$ range in \BH\ mass around the best fit relation. (b)
\label{fig:BHsig}\MBH-$\sigma$ correlation.  The dashed lines represent a $\pm
0.5$ range in \BH\ mass around the best fit relation. Data from \cite{kg}.}
\end{figure}
\section{\label{sec:BHgal}Black Holes and Host Galaxy Properties}

The many reliable \BH\ detections ($\sim 40$) allow demographical studies.
\cite{kr} reported a correlation between \BH\ mass and bulge luminosity which
was confirmed by following studies \cite{magorrian, ho99}.  More recently a
tighter correlation has been found between the \BH\ mass and the central
stellar velocity dispersion of the host spheroid \cite{ferrarese00,
ghebardt00b}. The slope of the $\MBH-\sigma$ correlation is still a matter of
debate and is in the range $\MBH\sim \sigma^{4-5}$. The two correlations are
displayed and compared in Fig.\ \ref{fig:BHsig}. The $\MBH-\sigma$ correlation
does appear tighter but this could be  a consequence of uncertainties in bulge
luminosity estimates.  Indeed, \cite{mclure01} estimate the \BH\ mass in a
sample of AGNs using the width of the H$\beta$ line and re-analyze the
\MBH-\Lbulge\ relation finding that its scatter is similar to that of the
\MBH-$\sigma$ relation ($\sim 0.3$dex). This comes from better estimates of the
bulge luminosities obtained from full 2D bulge-disk decomposition and from the
use of the R band which is less contaminated by extinction and star formation than the
commonly used B band.  The \BH\ mass also correlates with the light
concentration of the bulge and the tightness of the correlation
is comparable with that of the \MBH$-\sigma$ relation \cite{graham}. 

These correlations have three important consequences.  The first one is that
the \BH\ formation must be related to the formation of the host spheroid. 
The second one is that the \MBH\ values estimated using
reverberation mapping and H$\beta$ line widths are reliable since
they agree well with the \MBH$-\sigma$ relation. This also implies that
\BH\ masses of AGNs are on average indistinguishable
from those of normal galaxies \cite{ghebardt00c,mclure01}. 
Finally, the third consequence is that the correlations
are "cheap" empirical estimators of the 
\BH\ mass in large samples of objects.
For example, \cite{merritt00} use the \MBH$-\sigma$ relation to
estimate the \MBH/\Mbulge\ ratio in a sample
of elliptical galaxies. They find that $\log(\MBH/\Mbulge)\sim -2.90$
with r.m.s. $\sim 0.45$.  This implies that the local density in black holes is
$\rhobh\sim 5\xten{5}\Msun\MPC\3$.  

\section{\label{sec:agn&bh}Black Holes and AGN Activity}

\begin{figure}[t!]
\parbox[c]{0.48\linewidth}{
\epsfig{file=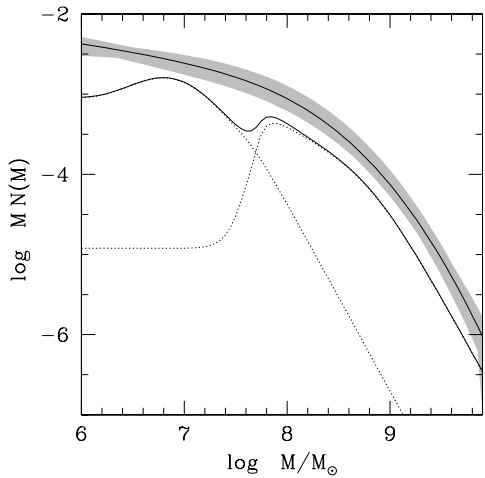, width=\linewidth}}
\hskip 0.06\linewidth
\parbox[c]{0.44\linewidth}{
\caption{\label{fig:BHMF}
Local \BH\ mass function $N(M)$ (\#/d$M$/\MPC$^{3}$) from bulges (solid line in
gray area) compared with that expected from the activity of type 1 AGNs.
The dotted lines
represent the contributions from low and high luminosity objects. 
The mass function of AGN relics 
should then be multiplied by $\sim 2-4$, the ratio of type 2 to type 1 objects,
to account for the whole AGN population. 
Adapted
from \cite{marconi01}.}}
\end{figure}

To test if the \BHs\ in the nuclei of nearby galaxies are relics of AGN activity
one
can estimate the integrated comoving energy density from AGNs
\begin{equation}
u = \int_0^\infty\int_0^\infty \Phi(L,z)L dL \frac{dt}{dz} dz = 1.06\xten{-15}
\ERG\CM\3
\end{equation}
$\Phi$ is the luminosity function of type 1 AGNs (e.g. quasars or, in general, AGNs with broad emission lines
in their optical spectra) used by \cite{marconi01}.
With an accretion efficiency $\epsilon$ the relic mass density is
$\rhobh = u/(\epsilon c^2) =
1.74\xten{5}(\epsilon/0.1)^{-1}\Msun\MPC\3$.
This number must be multiplied by the ratio between type 2 (i.e. those without broad emission lines) and
type 1 AGNs, $R_{21}$, to account for the whole AGN population. Hence $\rhobh \sim 7\xten{5}\Msun\MPC\3$ with
$R_{21}=4$, the canonical number used in AGN unified models.
This is in agreement
with $\rhobh = (5\pm
2)\xten{5}\Msun\MPC\3$, the \BH\ mass density estimated by combining the \MBH$-\Lbulge$ correlation with bulge luminosity functions
 \cite{marconi01}.
This argument has been presented in many papers following from the
work by Soltan \cite{soltan} and Chokshi \& Turner \cite{cht}.
The density in relic \BHs\ can also be estimated from the X-ray background
emission: assuming that the XRB bump in the 10-30 keV spectral range
constitutes the integrated emission from {\it all} AGNs, one derives the AGN energy density and then $\rhobh\simeq 3-6\xten{5}\Msun\MPC\3$ \cite{fabian99,salucci99}.
All the above values are in good agreement. In particular, the expected density in AGN relics matches the \BH\ mass density derived from local bulges
indicating that most of the
\BH\ masses are relics of past AGN activity.
It is possible to reproduce the above arguments in a more refined way dealing
directly with the \BH\ mass function and not only with integrated values (\rhobh). \cite{marconi01} compare the \BH\ mass function (MF) expected from AGN activity with the \BH\ MF derived from local bulges \cite{salucci99}.
The main conclusions are that, using standard assumptions on AGN activity,
compatible with current knowledge, one can reproduce the \BH\ MF of local bulges
{\it both} in shape and normalization (Fig.\ \ref{fig:BHMF}).

There have been suggestions of a correlation between radio emission and the
mass of the \BH. In particular it seems the radio-loud quasars are
characterized by the most massive \BHs\ (\cite{franceschini,mclure01}
and references therein).
However, Ho \cite{ho01} finds that the radio
continuum power, either from the whole galaxy or from the nuclear core alone,
correlates poorly with \MBH. The degree of
radio loudness (radio-to-optical luminosity) is strongly inversely correlated
with $L/\LEdd$, which is taken as evidence for advection-dominated accretion.
The issue of the correlation \MBH-Radio Power is still much debated and potentially
very important because it could lead to the much sought unification between radio-loud and radio-quiet AGNs.

\section{Black Holes and Galaxy formation}

The physical reasons behind the correlations between \MBH, bulge mass and $\sigma$
are currently being investigated by several authors and only a few examples will
be presented here.  \cite{silk98,fabian} propose a scenario in which \BH\
growth is self regulated: the \BH\ forms in the galaxy nucleus before all the
bulge gas is turned into stars. Then it accretes gas giving rise to
quasar-like activity.  When the \BH\ mass is large enough the radiation
pressure and wind produced by AGN activity will sweep away the gas,
blocking  growth and star formation in the spheroid. \cite{silk98} use
an energy argument to determine this critical mass of the \BH\ and find
$\MBH\propto \sigma^5$. However \cite{fabian} uses a different argument based
on force balance, finding $\MBH\propto \sigma^4$.  \cite{adams} model the
bulge as a rotating isothermal sphere and the \BH\ growth is stopped when the
centrifugal radius of the collapse flow exceeds the capture radius of the \BH,
implying $\MBH\propto \sigma^4$.  Similarly \cite{cavaliere} show that if the
\BH\ growth is completely  self regulated by the luminosity output, $\MBH\propto
\sigma^5$  is to hold at all $\sigma$'s. On the other hand, if the feedback is
not important $\MBH\propto \sigma^4$ will hold at high $\sigma$ while at lower
values it will soften to  $\MBH\propto \sigma^3$, when the growth is completely supply-limited.

An accurate determination of the slope of the $\MBH-\sigma$ correlation is thus
important to distinguish between self-regulated \BH\ growth ($\MBH\propto
\sigma^5$)  or growth determined by ambient conditions ($\MBH\propto
\sigma^4$).  In the former case, the bulge mass in stars is set by \MBH.

\cite{h&k} show that the observed correlations
\MBH-$\Lbulge$ and \MBH-$\sigma$ can be reproduced both in slope and scatter
with their model in which bulges and
supermassive black holes both form during major mergers.
Observational support to the idea of \BH\ from merging comes from 
\cite{ravind}. They study the central cusp slopes and core
parameters of early type galaxies using a sample of objects observed 
with HST and
find that
the observational trends are reproduced in the framework of binary black
hole mergers but not in that of adiabatic growth models.
\cite{ciotti} combine the \MBH-$\sigma$ relation with other
scaling relations for elliptical galaxies such as the Faber-Jackson relation
and the fundamental plane relation. In order not to produce effective
radii of elliptical galaxies larger than observed, the rule for adding the
mass of merging \BHs\ must be substantially different from what is assumed, or the
merging process must involve a significant dissipative phase.

\section{Conclusions}

The observational evidences presented so far suggest the ubiquity of \BHs\
in the nuclei of all bright galaxies, regardless of their activity.
\BH\ masses correlate with masses and luminosities of the
host spheroids and, more tightly, with stellar velocity dispersions.
These correlations constrain \BH\ formation and growth but can also
be used as empirical estimators of \BH\ masses in large samples
of objects.

Accretion of matter onto a \BH\ during AGN phases can reproduce the mass
function and account for the local density of \BHs\ thus implying a strict
relationship between \BH\ growth and AGN activity.  

However several issues remain to be solved since the field
"Massive Black Holes in Galactic Nuclei" is still young.
Following are some issues, both general and particular, which should be tackled in the near
future.\\
-- We have to prove unambiguously that the massive dark objects present in galactic nuclei are \BHs\ by detecting relativistic motions
close to their event horizon.\\
-- Is the $\MBH-\sigma$ correlation really tighter than the $\MBH-\Lbulge$
correlation or is it a consequence of inaccurate determinations of \Lbulge ?\\
-- What is the slope of the $\MBH-\sigma$ correlation? Solving this issue
will indicate if \BH\ growth is self-regulated or not.\\
-- Up to what redshift can the correlations be used to estimate \BH\ masses?
Or alternatively, are the $\MBH-\Lbulge$ and $\MBH-\sigma$ correlations valid
throughout the evolution of a galaxy or just during its final stages?\\
-- Some observational evidences and theoretical models suggest that a
massive black hole could form from the merging of smaller \BHs. What is the
importance of merging in the growth of a \BH\ related to mass accretion during
AGN phases?

Apart from HST, which will continue detecting \BHs\ in nearby galactic nuclei,
further developments can be expected from the use
of adaptive optics with 8m-class telescopes like VLT and Keck
and, of course, from the launch of NGST. 
The use of interferometric techniques in \BH\ searches must still
be assessed but could produce a real breakthrough in spatial resolution
allowing one to probe down to the milli-arcsecond level.

\vfill
{\noindent\bf Acknowledgments.} I am very grateful to the organizers for the
invitation and the financial support. I wish to thank Alessandro Capetti,
Witold Maciejewski and Neil Nagar for reading the manuscript and provinding
comments and suggestions.

\end{document}